\documentclass[twocolumn]{aastex701}

\usepackage{amsmath,amssymb,graphicx,xcolor,makecell}

\newcommand{\cii}{[C\,{\sc ii}]}

\begin{document}

\title{A Close Quasar Pair in a Massive Galaxy Merger at $z=5.7$}

\correspondingauthor{Minghao Yue}

\author[0000-0002-5367-8021]{Minghao Yue}
\email{myue@mit.edu}
\affiliation{Steward Observatory, University of Arizona, 933 N. Cherry Ave., Tucson, AZ 85721, USA}

\author[0000-0003-3310-0131]{Xiaohui Fan}
\affiliation{Steward Observatory, University of Arizona, 933 North Cherry Avenue, Tucson, AZ 85721, USA}
\email{}

\author[0000-0003-2895-6218]{Anna-Christina Eilers}
\email{eilers@mit.edu}
\affiliation{MIT Kavli Institute for Astrophysics and Space Research, 77 Massachusetts Ave., Cambridge, MA 02139, USA}

\author[0000-0002-7633-431X]{Feige Wang}
\affiliation{Department of Astronomy, University of Michigan, 1085 S. University Ave., Ann Arbor, MI 48109, USA}
\email{}

\author[0000-0001-5287-4242]{Jinyi Yang}
\affiliation{Department of Astronomy, University of Michigan, 1085 S. University Ave., Ann Arbor, MI 48109, USA}
\email{}

\author[0000-0003-3769-9559]{Robert A. Simcoe}
\email{simcoe@space.mit.edu}
\affiliation{MIT Kavli Institute for Astrophysics and Space Research, 77 Massachusetts Ave., Cambridge, MA 02139, USA}

\begin{abstract}
Close quasar pairs are rare products of galaxy mergers in which both supermassive black holes (SMBHs) are actively accreting, offering strong constraints on merger-driven active galactic nuclei (AGN) evolution. Identifying close quasar pairs at $z\gtrsim4$ is challenging due to the declining quasar number density in the early Universe. Here we report the confirmation of a close quasar pair at $z=5.7$, J2037--4537, utilizing high-resolution Atacama Large Millimeter/submillimeter Array (ALMA) observations. 
The quasar host galaxies exhibit tidal disturbed features in both the far-infrared continuum emission and the {\cii} line emission, ruling out the doubly-imaged lensed quasar scenario. The two quasar hosts are massive $(M_\text{dyn}\gtrsim10^{10}M_\odot)$ and star-forming (SFR $\gtrsim500 M_\odot~ \mathrm{yr^{-1}}$). The confirmation of J2037--4537 puts a lower limit on the quasar pair fraction at $5.5<z<6$, $F_\text{pair}>1.2\%$, which is much higher than the quasar pair fraction at $z\lesssim4$. J2037--4537 is expected to form a gravitationally-bound SMBH binary within $\lesssim2$ Gyr. The elevated quasar pair fraction at $z>5.5$, as indicated by J2037--4537, likely contributes to the high gravitational-wave background reported by recent Pulsar Timing Array experiments.

\end{abstract}



\section{Introduction}
The hierarchical structure formation of the Universe naturally produces mergers between galaxies. Previous studies have demonstrated that galaxy mergers are capable of triggering powerful quasar activities \citep[e.g.,][]{dm05,hopkins08}. In some rare cases, a galaxy merger can ignite the supermassive black holes (SMBHs) in both merging galaxies, producing a close pair of quasars \citep[e.g.,][]{capelo15,capelo17}. The statistics of close quasar pairs, such as the quasar pair fraction and the luminosity distributions, put unique constraints on the models of galaxy merger evolution and quasar triggering \citep[e.g.,][]{shen23,ps25}.

Previous surveys have found about a hundred close quasar pairs with projected separations $\Delta d<10$ kpc \citep[e.g.,][]{silverman20, tang2021, shen21, lemon20, lemon22, chen22, chen23, chenyc25}. Most of these close quasar pairs are at redshifts $z\lesssim4$, consistent with cosmological simulations suggesting that their abundance peaks at cosmic noon \citep{ps25}. The number of close quasar pairs drops quickly at $z\gtrsim4$ {\citep[e.g.,][]{mat18, kulkarni19, shen20}}, due to the rapidly declining quasar number densities in the early Universe. So far, only a handful of close quasar pairs and candidates have been identified at $z>4$ \citep[][]{yue21,yue23,mat24}.

The first candidate kpc-scale quasar pair reported at $z>5$ is J2037--4537 \citep[][]{yue21}. J2037--4537 consists of two objects separated by $1\farcs24$, and both objects exhibit spectra of a $z=5.7$ quasar. Assuming J2037--4537 is indeed a physical quasar pair, the two quasars would have high luminosities of $L_\text{bol}>10^{46}\text{ erg s}^{-1}$ and a small projected separation of $\Delta d=7.4\text{ kpc}$. These features would make J2037--4537 a highly unique object in the early Universe. 

Nevertheless, \cite{yue21} could not fully determine the nature of J2037--4537, as this system might also be a doubly-imaged gravitationally lensed quasar. The decisive way to confirm a physical close quasar pair is to characterize their host galaxy emissions \citep[e.g.,][]{chen23}. Specifically, close quasar pairs exhibit tidal-disrupted features due to the merging event of their host galaxy, while the host galaxies of lensed quasars appear as lensed arcs.

In this paper, we present high-resolution Atacama Large Millimeter/submillimeter Array (ALMA) observations for J2037--4537. The ALMA observation reveals the quasar host galaxy emission in rest-frame far-infrared (FIR) wavelengths, allowing us to conclusively determine the nature of J2037--4537.
This paper is organized as follows. Section \ref{sec:data} describes the previous observations and the ALMA observation for J2037--4537. Section \ref{sec:method} presents the evidence supporting the physical quasar pair hypothesis and provides the FIR properties of the quasar host galaxies. We discuss the implication of J2037--4537 in Section \ref{sec:discussion} and summarize our finding in Section \ref{sec:summary}. Throughout this paper, we assume a flat $\Lambda$CDM cosmology with $\Omega_m=0.3$ and $H_0=70 \text{ km s}^{-1}\text{Mpc}^{-1}$.

\section{Data} \label{sec:data}

\subsection{The Target} \label{sec:target}
J2037--4537 was discovered in the Dark Energy Survey \citep[DES; e.g.,][]{desdr2} footprint. \cite{yue21} presented the
Hubble Space Telescope Advanced Camera for Surveys Wide Field Camera F850LP band image, as well as the optical-to-near-infrared spectra taken by the Low Dispersion Survey Spectrograph and the Folded-port InfraRed Echelle on the Magellan telescopes. These previous observations show that
J2037--4537 consists of two point sources separated by $1.24''$, both exhibiting the spectrum of a type-I quasar at redshift $z=5.7$. No other objects are detected between the two point sources, and the spectral shapes of the two quasars are significantly different. Accordingly, \cite{yue21} argued that J2037--4537 is a promising close quasar pair candidate and is not likely a doubly-imaged lensed quasar. {Based on the physical quasar pair scenario, \citet{yue21} measured the bolometric luminosities of the two quasars to be $\log L_\text{bol}[\text{erg s}^{-1}]=47.1$ and $46.9$, and estimated their SMBH masses to be $\log M_\text{BH}[M_\odot]=8.60$ and 8.45, respectively.}

Nevertheless, the above-mentioned observations cannot completely rule out the strong lensing hypothesis. Specifically, the non-detection of the lensing galaxy in the F850LP band can be explained by a red lensing galaxy at $z\gtrsim3$, and the significantly different spectral shapes of the two quasar images might be a result of microlensing and differential reddening.
To decisively confirm J2037--4537 as a physical quasar pair, we need to detect the host galaxy emissions and characterize the tidal-disrupted features of the merging quasar hosts. ALMA is the ideal facility for this experiment, as the rest-frame FIR emission of quasars is dominated by their host galaxies.

\subsection{ALMA Observations} \label{sec:obs}

\begin{figure*}
\centering
\includegraphics[width=0.4\linewidth]{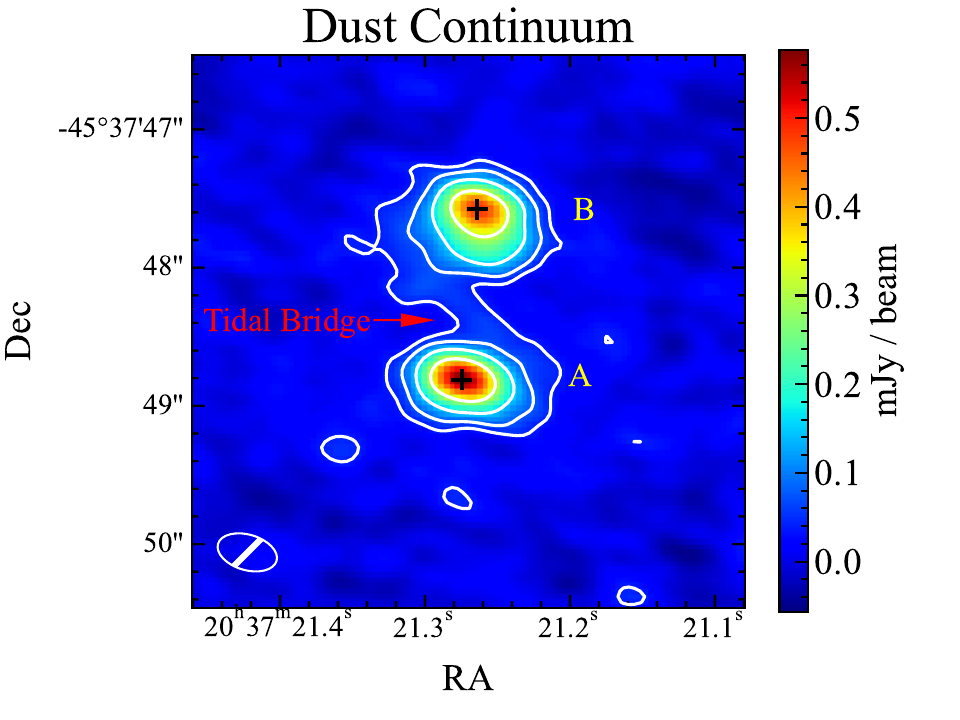}
\includegraphics[width=0.4\linewidth]{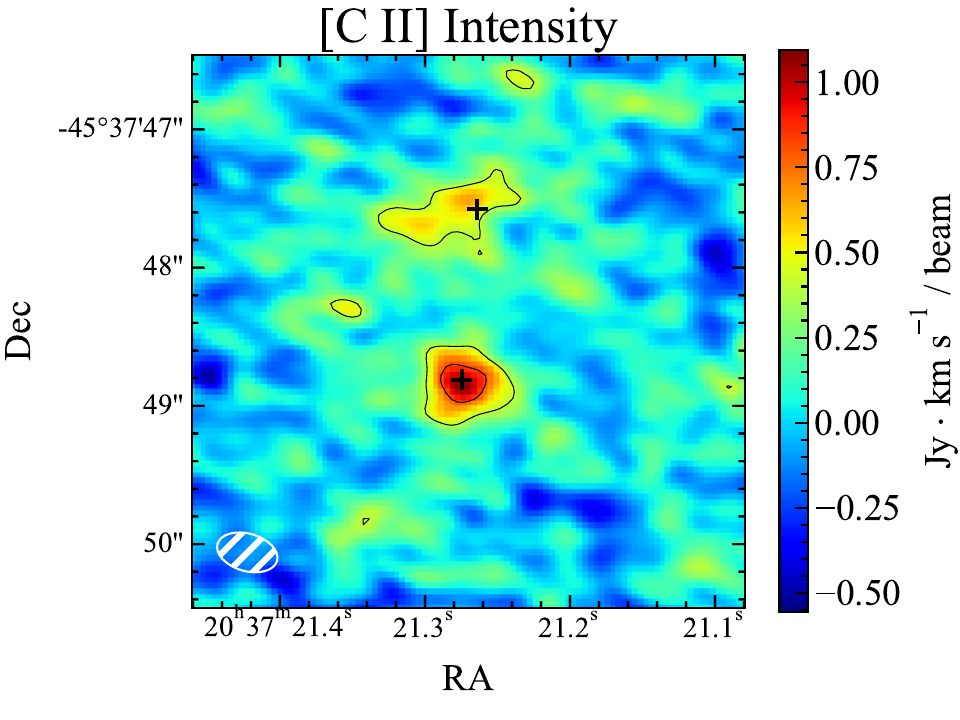}
\includegraphics[width=0.4\linewidth]{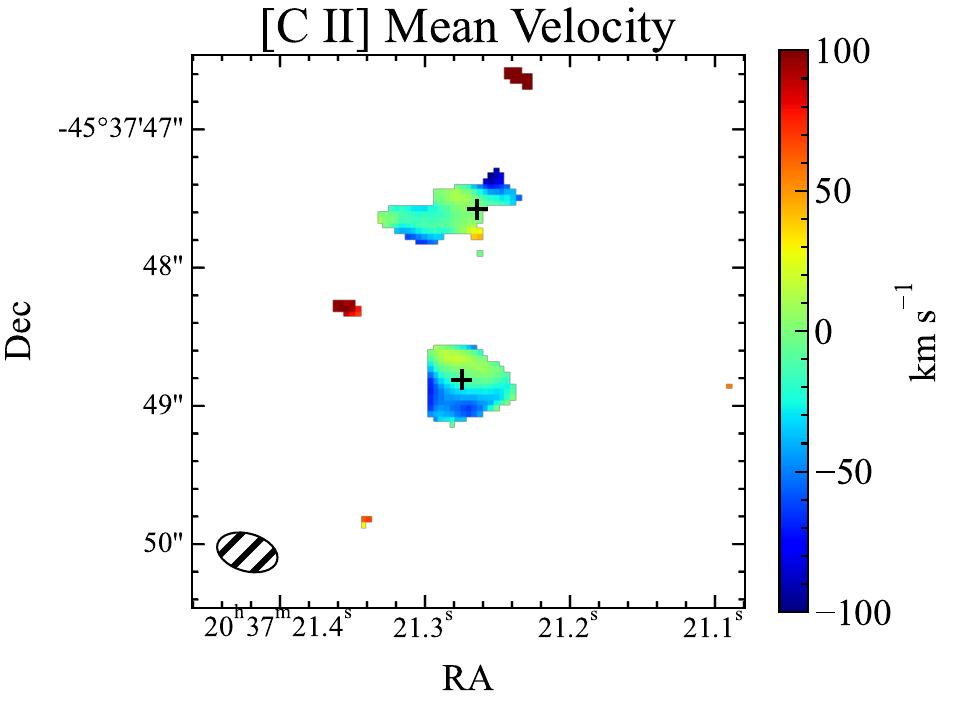}
\includegraphics[width=0.4\linewidth]{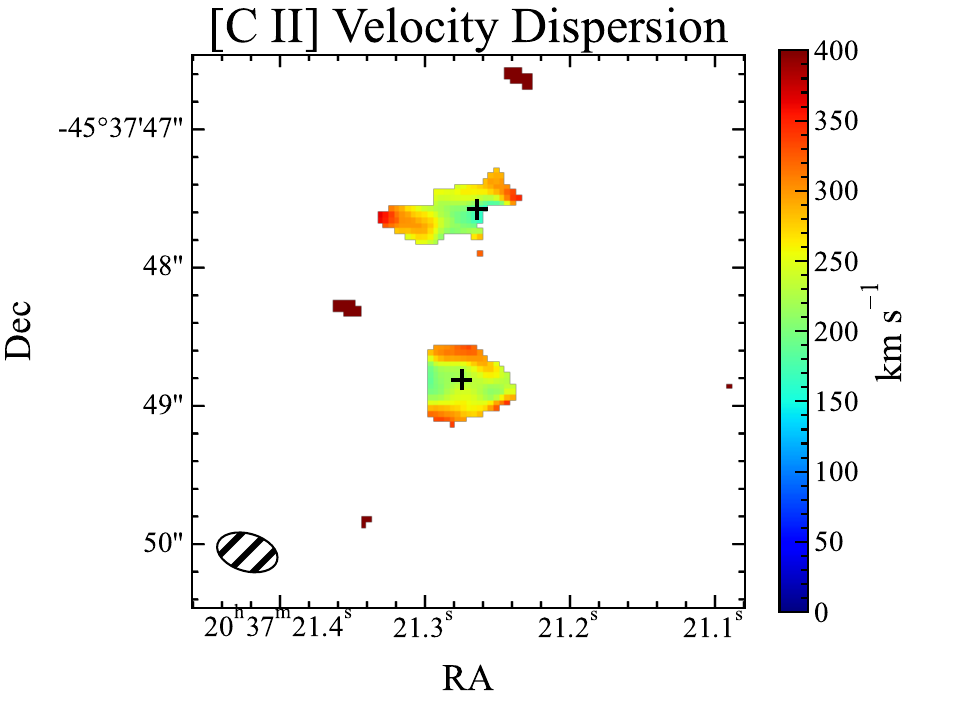}
\caption{
The ALMA observation of J2037--4537. The images are cleaned with natural weighting. The cleaned beam has sizes of $0\farcs44\times0\farcs26$, demonstrated by the hatched ellipses. The black crosses mark the optical positions of the quasars as measured by previous {\em HST} imaging \cite{yue21}.
{\em Upper Left:} The FIR continuum. The white lines mark $2\sigma,4\sigma,8\sigma,$ and $16\sigma$ contours.
{We denote the southern object as Quasar A and the northern one as Quasar B.
{\em Upper Right:} The integrated {\cii} intensity (the moment 0 map). The black lines mark $2.5\sigma$ and $5\sigma$ contours. 
The continuum exhibit tidal bridges between the two quasars, confirming that J2037--4537 is a physical quasar pair residing in an on-going galaxy merger.
{\em Lower Left:} The {\cii} velocity field. 
{\em Lower Right:} The {\cii} velocity dispersion map. 
For the lower panels, we only show pixels with $\text{S/N}>2.5$ in the {\cii} moment 0 map. Quasar B exhibits irregular structures indicative of tidal disruption.
The velocity gradient for both quasars are much smaller than their velocity dispersions, indicating that both quasars are dispersion dominated.
}}\label{fig:alma}
\end{figure*}

J2037--4537 was observed by ALMA in Band 7 under configuration C43-5 in August 2022 (Program ID 2021.1.01052.S). We tuned the four 1.875 GHz-wide spectral windows (SPWs) at 284.451 GHz, 286.263 GHz, 296.450 GHz, and 298.263 GHz. The on-source exposure time is 5,564 seconds.
We reduce the ALMA data using the Common Astronomy
Software Applications \citep[CASA;][]{casa} version 6.5.4.9. 
To measure the continuum emission, 
we obtain the continuum visibility by splitting out the line-free channels (i.e., with $\nu_\text{obs}>284.5$ GHz).
{To obtain the cleaned continuum image, we run task \texttt{tclean} interactively using natural weighting, until the residual shows no significant patterns beyond the background noise level. The cleaned continuum image is shown in Figure \ref{fig:alma}}; the clean beam size is $0\farcs44\times0\farcs26$. {The continuum emission map exhibits two objects with a tidal bridge connecting them, and the tidal tail is detected at a significance of $>2\sigma$. In the rest of this paper, we refer the southern object as Quasar A and the northern object as Quasar B.}

\begin{figure}[h]
\centering
\includegraphics[width=1\linewidth]{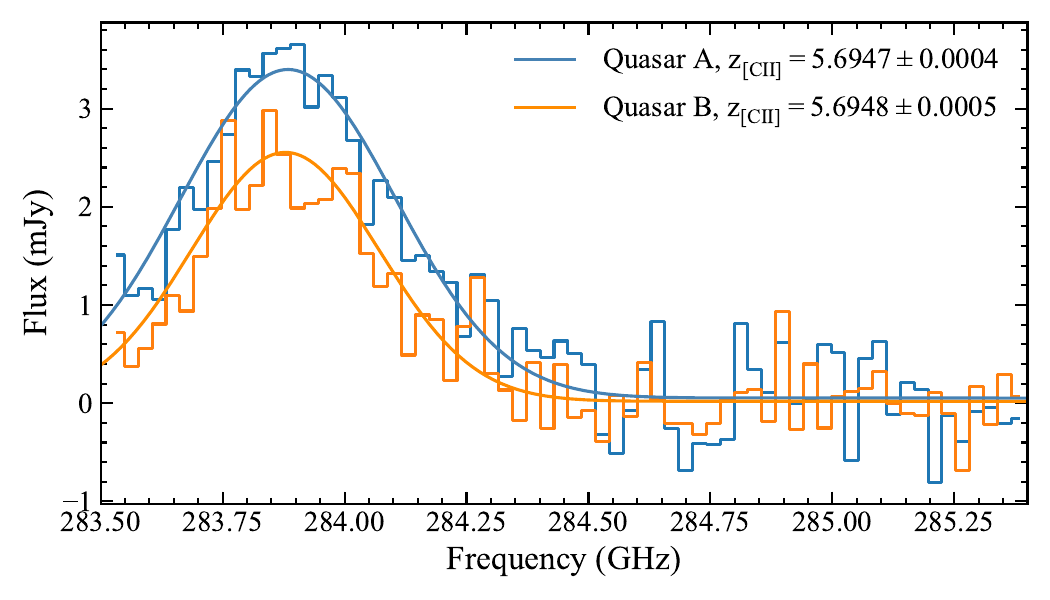}
\caption{The extracted {\cii} spectra of the two quasars (step plots) and the best-fit Gaussian profile for the lines (solid lines). The two quasars have {\cii} redshifts of $z_\text{\cii}=5.69$ and do not show a significant line-of-sight velocity difference. We note that the {\cii} lines fall on the edge of the spectral windows, and the continuum with frequency lower than {283.50 GHz} is not covered.}
\label{fig:line}
\end{figure}

To measure the {\cii} line emission,
we use CASA task \texttt{uvcontsub} to fit a linear model to the line-free channels 
and subtract the linear model from the total visibility to obtain the line-only visibility.
We then run \texttt{tclean} {interactively} using natural weighting to get the {\cii} line cube.
We extract the {\cii} spectra of the two quasars using circular apertures centered on the brightest pixels of each quasar in the continuum cleaned image.  The diameter of the apertures is $0\farcs8$, sufficient to cover most emission from both quasars. Figure \ref{fig:line} shows the extracted spectra; the {\cii} line of the two quasars falls at the edge of the spectral coverage. For each quasar, we fit the spectrum around the {\cii} lines as a Gaussian profile plus a constant using the Markov Chain Monte Carlo (MCMC) method. The resulting redshifts are $z_{\rm A}=5.6947 \pm 0.0004$ and $z_{\rm B}=5.6948 \pm 0.0005$ for quasar A and quasar B, respectively.

We produce the {\cii} moment 0 map using task {\texttt{immoments}} with a frequency range $283.534\leq\nu_\text{obs}[\text{GHz}]\leq284.301$, which is symmetrically positioned on the {\cii} line center. For the mean velocity and the velocity dispersion maps, we fit each pixel in the cleaned {\cii} cube as a Gaussian profile plus a constant, and take the mean and standard deviation of the Gaussian profile as the mean velocity and velocity dispersion at that pixel. {Some examples of this pixel-by-pixel spectral fitting are shown in Appendix \ref{sec:app}.} The Upper Right Panel and the Lower Panels in Figure \ref{fig:alma} show the cleaned {\cii} moment 0 map, the mean velocity map, and the velocity dispersion map. {In the moment 0 map, quasar B shows irregular structures indicative of tidal disruption.}

\section{Properties of J2037--4537} \label{sec:method}

\subsection{Confirming the Close Quasar Pair Hypothesis} \label{sec:lensmodel}


{The tidal bridge in the dust continuum map and the irregular structures of quasar B in the {\cii} moment 0 map indicate that J2037--4537 is an on-going galaxy merger.} Additionally, we do not see features like lensed arcs in the clean images. These results essentially confirm that J2037--4537 is a physical close quasar pair.

\begin{figure*}
    \centering
    \includegraphics[width=1\linewidth]{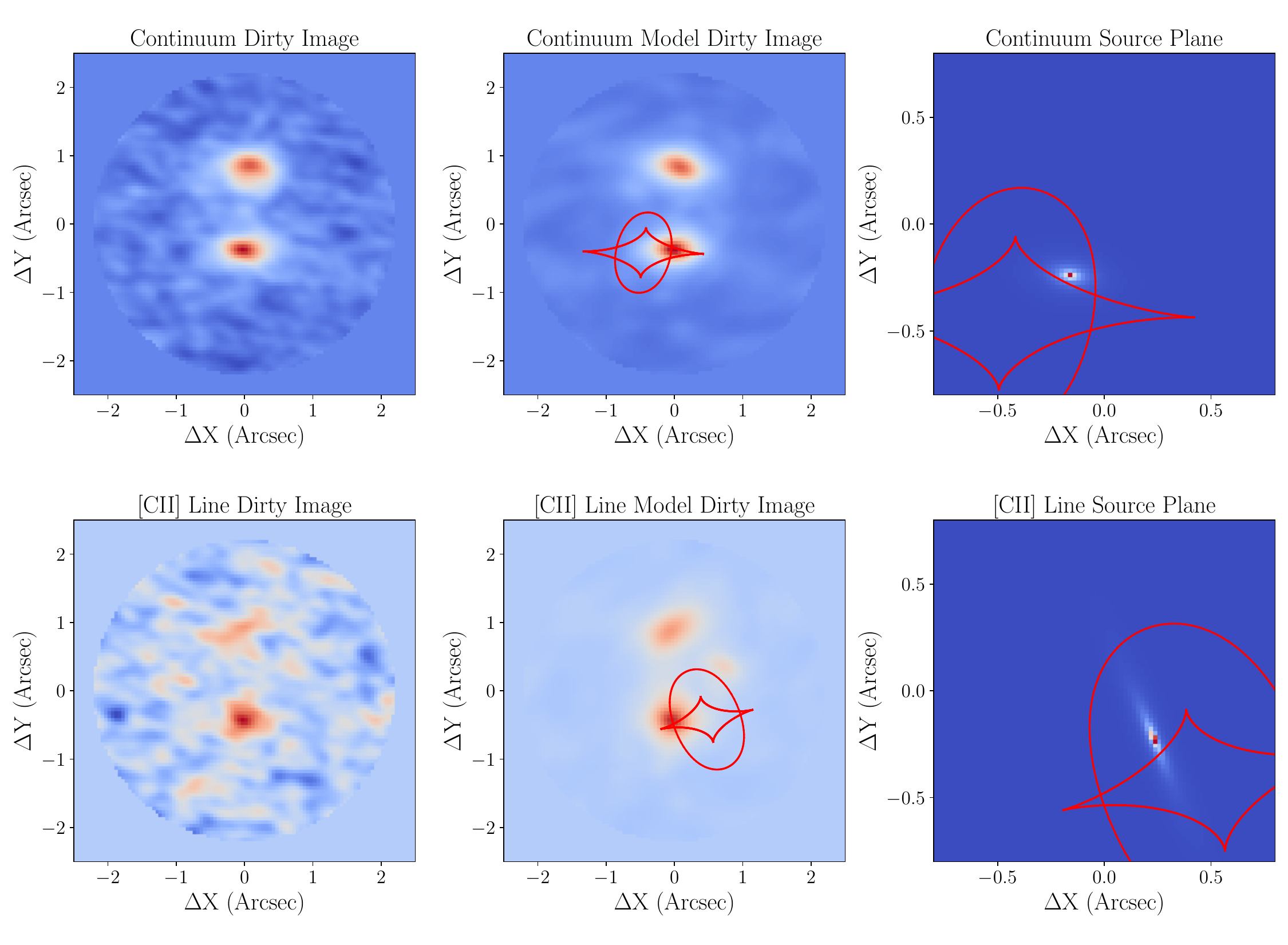}
    \caption{The best-fit lensing model for the continuum and {\cii} line emission. From left to right: the observed dirty image, the modeled dirty image, and the source plane emission. The red lines mark the caustics of the best-fit lensing models. The best-fit lensing configuration for the continuum emission and the {\cii} line differ significantly, essentially ruling out the strong lensing hypothesis for J2037--4537.}
    \label{fig:lensingmodel}
\end{figure*}

To securely rule out the strong lensing hypothesis, we attempt to fit the ALMA observations using lensing models.
We assume a singular isothermal elliptical density profile for the potential lensing galaxy, and add an external shear component to account for perturbations from other galaxies near the line-of-sight.
We use S\'ersic profiles to describe the source plane continuum and velocity-integrated {\cii} line emission.
We use \texttt{PyAutoLens} \footnote{https://github.com/Jammy2211/PyAutoLens} \citep{pyautolens, Nightingale2015, Nightingale2018}  to perform lens modeling. 
The results are shown in Figure \ref{fig:lensingmodel}. The best-fit lensing models for the continuum and the {\cii} line have significantly different deflector galaxy positions and Einstein radii, as demonstrated by the lensing caustics. Furthermore, the best-fit models suggest extremely small sizes for the quasar host galaxy; the half-light radius is $0.5$ kpc for the continuum emission and $0.7$ kpc for the {\cii} line. In contrast, quasar host galaxies at $z\gtrsim6$ have radii of $\gtrsim1$ kpc \citep[e.g.,][]{decarli18,venemans20}. We also notice that the best-fit {\cii} model has an unphysically small axis ratio $(q=0.17)$. All these features indicate that J2037--4537 cannot be well-described by a lensing model.

\subsection{Quasar Host Galaxy Properties} \label{sec:properties}

After confirming J2037--4537 as a physical quasar pair residing in an ongoing galaxy merger, we can evaluate the properties of the two merging quasar host galaxies. 
We first measure their continuum fluxes using the CASA task {\texttt{imstat}}, applying the same apertures used for extracting the {\cii} spectra (Section \ref{sec:obs}). We estimate the flux error using an annulus with an inner radius of $6''$ and an outer radius of $12''$, centered at the phase center. We also measure the sizes of the quasar host galaxies using CASA task {\texttt{imfit}}, which fits the emission as 2D Gaussian profiles. 
We measure the {\cii} line fluxes and full-width half maximum (FWHM) for the two quasars using the Gaussian fit in Figure \ref{fig:line}, and measure the sizes of the {\cii} emitting regions by running {\texttt{imfit}} on the {\cii} moment 0 map.


Using the continuum flux, we model the FIR spectral energy distribution (SED) as a modified black body \cite[e.g.,][]{decarli18,Mazzucchelli25}:
\begin{equation}
    F_\text{obs}(\nu)= \frac{f_\text{CMB} (1+z)}{D_L^2} \kappa_{\nu, \text{rest}}(\beta)M_\text{dust}B_{\nu,\text{rest}}(T_\text{dust})
\end{equation}
where $B_\nu(T)$ is the black body spectrum, $f_\text{CMB}=1-B_\nu(T_{\text{CMB},z})/B_\nu(T_{\text{dust},z})$ is the correction factor for the cosmic microwave background, $\kappa_{\nu, \text{rest}}(\beta)=0.077\left(\frac{\nu_\text{rest}}{352\text{GHz}}\right)^{\beta}$ is the dust mass opacity coefficient, $M_\text{dust}$ is the dust mass, and $D_L$ is the luminosity distance.
We adopt a canonical dust temperature of $T_\text{dust}=47$K and emissivity of $\beta=1.6$ \citep{beelen06}.
We then evaluate the FIR (42.5-122.5 {\micron}) and total-infrared (TIR, 8-1000 {\micron}) luminosities of the quasar host galaxies by integrating the modeled SEDs.
Based on the TIR luminosity, we estimate the star formation rates following the relation in \cite{kennicutt12}:
\begin{equation}
    \frac{\text{SFR}_\text{TIR}}{M_\odot\text{ yr}^{-1}}=1.49\times10^{-10} \times\frac{ L_\text{TIR}} {L_\odot}
\end{equation}
{We note that the the assumptions of $T_\text{dust}=47$K and $\beta=1.6$ introduce systematic uncertainties in $L_\text{TIR}$ and $\text{SFR}_\text{TIR}$. At fixed observed sub-mm continuum flux, higher $T_\text{dust}$ and larger $\beta$ lead to higher $L_\text{TIR}$ \citep[e.g.,][]{beelen06}. Recent observations have found a wide range of $34\lesssim T_d~{\rm [K]}\lesssim65$ and $1.0\lesssim \beta\lesssim2.5$ \citep[][]{costa26}. Future multi-band FIR photometry is needed to directly constrain $T_\text{dust}$ and $\beta$ for J2037--4537.}

We compute the {\cii} luminosity following the relation in \cite{decarli18} and \cite{wang24}:

\begin{equation}
    \frac{L_\text{\cii}}{L_\odot} = 1.04\times10^{-3} \times \frac{F_\text{\cii}}{\text{Jy km s}^{-1}} \frac{\nu_\text{obs}}{\text{GHz}} \left(\frac{D_L}{\text{Mpc}}\right)^2
\end{equation}
where $\nu_\text{obs}$ is the observed frequency of the {\cii} line.

Since the two quasar hosts are dispersion-dominated with no visible ordered rotation (Figure \ref{fig:alma}),
we estimate the dynamical masses of the quasar hosts using the relation in \cite{decarli18}, which is suitable for dispersion-dominated systems:

\begin{equation}
    M_\text{dyn}=1.5 R_\text{maj} \sigma^2 / G
\end{equation}
where $R_\text{maj}$ is the semi-major axis of the quasar host as measured by CASA task \texttt{imfit}.

Table 1 summarizes all the measured properties. The two quasar host galaxies are massive (with $M_\text{dyn}\gtrsim10^{10}M_\odot$) and actively forming stars (with $\text{SFR}\gtrsim500M_\odot\text{ yr}^{-1}$). The SMBH masses of the two quasars are $M_\text{BH,A}=3.9\times10^8M_\odot$ and $M_\text{BH,B}=2.8\times10^8M_\odot$, respectively \citep[][]{yue21}, and the $M_\text{BH}/M_\text{host}$ ratio is about {$1\%$}. These numbers are similar to other $z\sim6$ quasars previously reported by ALMA \citep[e.g.,][]{decarli18,venemans20,izumi19}.

\begin{deluxetable}{ccc}
\tablenum{1}
\label{tbl:properties}
\tablecaption{Far-infrared Properties of J2037--4537}
\tablewidth{0pt}
\tablehead{\colhead{Object} & \colhead{A} & \colhead{B}}
\startdata
\hline
R.A.   &  20:37:21.27   &  20:37:21.26  \\
Dec    & -45:37:48.8   & -45:37:47.5  \\\hline
$z_\text{\cii}$  & {$5.6947 \pm 0.0004$}   & {$5.6948 \pm 0.0005$}  \\
$F_\text{287GHz}$ (mJy)    & $0.807\pm0.008$   & $0.828\pm0.007$  \\
$F_\text{{\cii}}~(\text{Jy} \cdot \text{km s}^{-1})$    & {$1.95\pm0.26$}   & {$1.30\pm0.18$}\\
$L_\text{FIR}~(L_\odot)$    & $(3.22\pm0.14)\times10^{12}$   & $(2.60\pm0.09)\times10^{12}$  \\
$L_\text{TIR}~(L_\odot)$    & $(4.52\pm0.19)\times10^{12}$   & $(3.66\pm0.13)\times10^{12}$  \\
$L_\text{{\cii}}~(L_\odot)$    & {$(1.70\pm0.22)\times10^9$}  & {$(1.13\pm0.15)\times10^9$}  \\
$\text{SFR}_\text{TIR}~(M_\odot\text{ yr}^{-1})$    & $672\pm29$   & $543\pm19$  \\
\hline
Deconvolved Size, continuum \tablenotemark{1} $('')$ & $0.31\times0.22$   & $0.37\times0.32$  \\
Deconvolved Size, {\cii} \tablenotemark{1} $('')$ & {$0.40\times0.28$}   & {$1.08\times0.33$}  \\
$R_\text{major, continuum}$ (kpc) & $0.90\pm0.14$ & $1.08\pm0.19$ \\
$R_\text{major, \cii}$ (kpc) & {$1.16\pm0.20$} & {$3.16\pm0.65$} \\
{\cii} line FWHM $(\text{km s}^{-1})$ & {$549\pm79$} & {$479\pm75$} \\
$M_\text{dyn} ~(M_\odot)$    & {$(2.2\pm0.7)\times10^{10}$}   & {$(4.6\pm1.7)\times10^{10}$}  \\\hline
\enddata
\tablenotetext{1}{The full-width half maximum of the deconvolved emission, from CASA task \texttt{imfit}.}
\end{deluxetable}

\section{Discussions} \label{sec:discussion}

\subsection{Quasar Pair Fraction at $z>5.5$} \label{sec:pairfrac}

One critical property about close quasar pairs is the quasar pair fraction $(F_\text{pair})$, usually defined as the number of quasar pairs with separation $\Delta d<30$ kpc divided by the total number of quasars \citep[e.g.,][]{dr19}. Previous observations have found a quasar pair fraction of $\sim10^{-3}$, which decreases with luminosity and has little redshift evolution at $1\lesssim z\lesssim4$ \citep[e.g.,][]{silverman20, shen23}. Meanwhile, cosmological simulations have investigated the pair fractions of faint AGNs $(L_\text{bol}\lesssim10^{43}\text{ erg s}^{-1})$, finding that $F_\text{pair}$ decreases with luminosity and might have a subtle increase towards $z\gtrsim5$ \citep[][]{dr19,ps25}.

The confirmation of J2037--4537 allows us to constrain the quasar pair fraction at $z>5.5$. To do this, we estimate the number of quasars at $5.5<z<6$ in the DES field, where J2037--4537 was discovered. The DES has a footprint of 5,000 deg$^2$, and the survey depth in \cite{yue21} is 21 magnitude in the $z$-band. Using the quasar luminosity function in \cite{mat18} and the template quasar spectrum in \citet{vb01}, we estimate the number of $5.5<z<6$ quasars with $m_z<21$ in the DES area to be $N_\text{qso}=83$. We thus infer a pair fraction of $F_\text{pair}>1/83=1.2\%$. This value is a lower limit because the survey of \citet{yue21} is incomplete. 

By assuming that J2037--4537 is the only $z>5.5$ close quasar pair in the DES field, we can further evaluate the confidence interval of the derived $F_\text{pair}$.
We use the Poisson statistics for the number of quasar pairs, and the probability of finding $N$ quasar pairs can be expressed as
\begin{equation}
    P(N|\lambda)=\frac{\lambda^N e^{-\lambda}}{N!}
\end{equation}
where $\lambda=N_\text{qso}\times F_\text{pair}$ is the expected number of quasar pairs. 
The discovery of J2037--4537 indicates $N\geq1$. 
Following the analysis \cite{gelman2013bayesian}, the posterior distribution of $\lambda$ can be expressed by a Gamma distribution:
\begin{equation}
    P(\lambda|N)=\text{Gamma}(\lambda, \alpha+N, \beta+1)
\end{equation}
where $\alpha$ and $\beta$ determines the prior of $\lambda$. We take $\alpha=1$ and $\beta=0$, which corresponds to a flat prior.
By taking $N=1$, we evaluate that the probability of having $F_\text{pair}>4.28\times10^{-3}$ is $95\%$.

\begin{figure}
    \centering
    \includegraphics[width=1\linewidth]{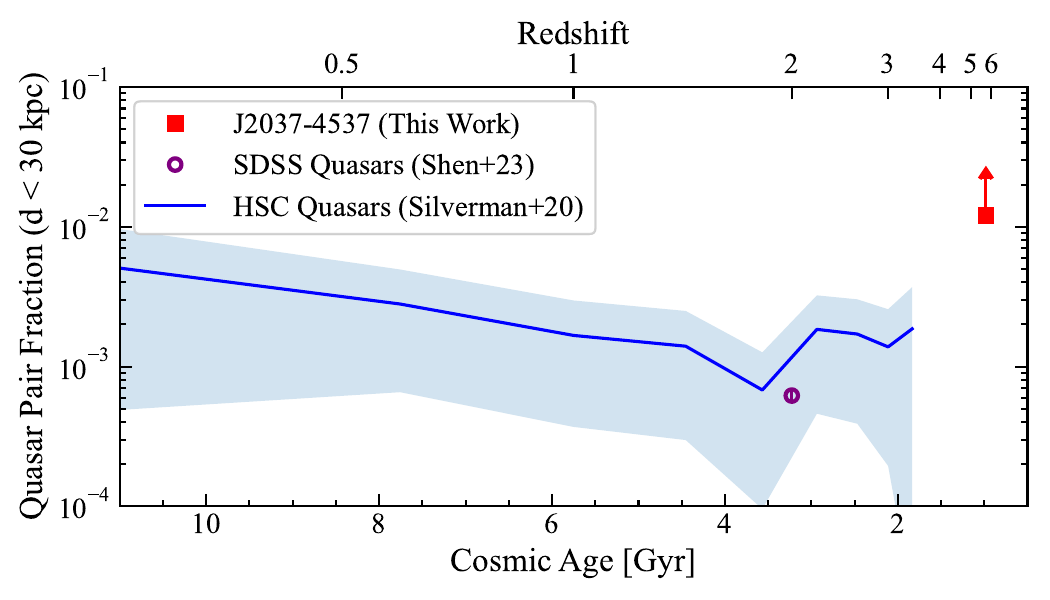}
    \caption{The redshift evolution of quasar pair fraction $(F_\text{pair})$, defined as the number of quasar pairs with separation $d<30$ kpc divided by the total number of quasars. At $z\lesssim3.5$, the quasar pair fraction is $F_\text{pair}\sim(0.5-2)\times10^{-3}$ with little redshift evolution. In comparison, J2037--4537 itself indicates $F_\text{pair}>1.2\%$ at $5.5<z<6$.}
    \label{fig:pairfrac}
\end{figure}

{The estimated quasar pair fraction at $z>5.5$ is much higher than that at $z\lesssim4$ \citep[e.g.,][see Figure \ref{fig:pairfrac}]{silverman20,shen23}. Specifically, \citet{silverman20} evaluated $F_\text{pair}$ at $0<z<3.5$ using 421 quasar pair candidates in the Hyper Suprime-Cam (HSC) Subaru Strategic Program \citep[][]{hsc}, which have $10^{44}\text{erg s}^{-1}<L_\text{bol}<10^{47.5}\text{erg s}^{-1}$; \citet{shen23} used 60 quasar pair candidates from Sloan Digital Sky Survey (SDSS) DR16 quasar catalog \citep[][]{lyke20} with $L_\text{bol}>10^{45.8}\text{erg s}^{-1}$. In comparison, both quasars in J2037--4537 have $L_\text{bol}\gtrsim10^{47}\text{erg s}^{-1}$ \citep{yue21}, comparable to the brightest quasar pairs in \citet{silverman20} and \citet{shen23}. Since AGN pair fraction decreases with AGN luminosity \citep[e.g.,][]{dr19,shen23,ps25}, the excess of $F_\text{pair}$ at $z\gtrsim5.5$ relative to $z\lesssim4$ might be more significant than Figure \ref{fig:pairfrac} suggests.} This result is in line with recent discoveries of other $z\gtrsim5$ quasar pairs \citep[e.g.,][]{yue23, mat24, izumi24} and dual AGNs \cite[e.g.,][]{liq25}. The elevated quasar pair fraction at $z\gtrsim5.5$ indicates that the merger-driven quasar triggering mechanism might be significantly different in the early Universe.

We note that current simulations do not have sufficient volumes to produce a pair of luminous quasars $(L_\text{bol}\gtrsim 10^{46}\text{ erg s}^{-1})$ at $z\gtrsim5$, and a direct comparison between the simulated and observed quasar pair fraction at high redshifts is not practical.


\subsection{Merger Timescale Estimate} \label{sec:merger}

Systems like J2037--4537 are likely progenitors of gravitationally-bound binary SMBHs that emit low-frequency gravitational waves (GWs).  We estimate the timescale for this process as follows.
In the first step, the two quasar hosts merge into a single galaxy. Assuming singular isothermal sphere (SIS) density profiles for the quasar host galaxy density profiles and circular orbits, the timescale of this step can be estimated by the Chandrasekhar dynamical friction timescale \citep[][]{Chandrasekhar43}. We follow the expression in \cite{chen23}:
\begin{equation}
    t^\text{gal}_\text{fric} = \frac{2.7}{\ln \Lambda} \frac{r}{30\text{ kpc}} \left(\frac{\sigma_\text{H}}{200\text{ km s}^{-1}}\right)^{2} \left(\frac{\sigma_\text{S}}{100\text{ km s}^{-1}}\right)^{-3} \text{Gyr}
\end{equation}
where $r$ is the distance between the two galaxies, $\ln \Lambda$ is the Coulomb logarithm, 
$\sigma_\text{H}$ is the velocity dispersion of the more massive progenitor galaxy, and $\sigma_\text{S}$ is the velocity dispersion of the less massive progenitor galaxy. Here we take $\ln \Lambda=\ln (1+M_1/M_2)=0.93$, which is a good approximation for major mergers \citep[e.g.,][]{jiang08,Solanes18}. We assume random orientations to evaluate the physical distance $r$, which gives $r=7.3/\cos(30^\circ)=8.4$ kpc. Plugging in these numbers yields $t^\text{gal}_\text{fric}=0.13$ Gyrs.

In the second step, the two SMBHs experience dynamical friction and sink into the center of the merged galaxy. Assuming an SIS density profile for the merged galaxy, the timescale of this step can be estimated as \citep[e.g.,][]{chen23}:
\begin{equation}
    t^\text{BH}_\text{fric} = \frac{19}{\ln \Lambda} \left(\frac{r}{5\text{ kpc}} \right)^2\left(\frac{\sigma_\text{H}}{200\text{ km s}^{-1}}\right)\left(\frac{10^8M_\odot}{M_\text{BH}}\right) \text{Gyr}
\end{equation}
where the typical value for the Coulomb logarithm is $\ln \Lambda\sim6$. By assuming an initial separation of $r=8.3$ kpc for the two SMBHs, we estimate $t^\text{BH}_\text{fric}=2.0$ Gyr. We note that the actual timescale is likely shorter, as the separation of the two SMBHs will decrease after the two quasar hosts merge. 

The total timescale for J2037--4537 to form a gravitationally-bound binary SMBH is $t_\text{total}=t^\text{gal}_\text{fric}+t^\text{BH}_\text{fric}\approx2.1$ Gyrs. In other words, J2037--4537 will form a binary SMBH at $z\sim2$. Such binary SMBHs can emit low-frequency GWs detectable by pulsar timing array (PTA) observations. 
Recent PTA observations have found that the low-frequency GW background is higher than the predictions of galaxy evolution models \citep[e.g.,][]{iv22,nanograv,epta_inpta,chen25}; 
for example, the SMBH merger rate measured by the European Pulsar Timing Array and Indian Pulsar Timing Array collaborations \citep{epta_inpta} is about 1 dex higher than the simulation in \cite{chen19pta}.
The discovery of J2037--4537 indicates that SMBH pairs formed by $z\gtrsim5$ mergers might be more abundant than predicted by previous simulations,
potentially contributing to the observed high GW background.
Therefore, understanding the statistics of kpc-scale AGN pairs at $z\gtrsim5$ will be critical to interpreting the PTA observations.

\subsection{{Comparison to the $z=6.05$ Close Quasar Pair}}

{J2037--4537 are among the two close quasar pairs at $z>5$ confirmed so far. The other system was reported by \citet{mat24} and \citet{izumi24}, which consists of two $z=6.05$ quasars, namely HSC J121503.42--014858.7 and HSC J121503.55--014859.3. In the following text, we refer to this $z=6.05$ quasar pair as J1215--0148, and compare the properties of J2037--4537 to those of J1215--0148.}

{The two quasars in J1215--0148 have $L_\text{bol}\lesssim6\times10^{45}\text{ erg s}^{-1}$, located at the faint end of the quasar luminosity function. The two quasars have a projected separation of 12 kpc. ALMA observations for J1215--0148 identified a tidal bridge between the two quasars and a tidal tail extending from the brighter quasar, visible in both the far-IR continuum and the {\cii} line. The two quasar host galaxies have dynamical masses of $\sim9\times10^{10}M_\odot$ and $\sim5\times10^{10}M_\odot$. The FIR-based SFR of the two quasar hosts are $89\pm7~M_\odot \mathrm{yr}^{-1}$ and $<12~M_\odot \mathrm{yr}^{-1}$, respectively.} 

{J1215--0148 shares some properties with J2037--4537, including the tidal bridge between the two quasar hosts and the high mass ratio of the two quasar hosts ($M_1/M_2\gtrsim0.5)$. The key difference between the two systems is that J2037--4537 has higher luminosities, higher SFRs, and a smaller projected separation. \citet{izumi24} argued that J1215--0148 will have higher luminosities and SFRs when the galaxy merger enters the final coalesce phase. 
Therefore, J1215--0148 might evolve into a system similar to J2037--4537 at later times.}

\section{Summary} \label{sec:summary}

We confirm J2037--4537 as a kpc-scale quasar pair at $z=5.7$ using ALMA observations. The dust continuum and {\cii} line emissions clearly reveal the tidal bridge between the two quasars. We show that the dust continuum and {\cii} line cannot be described by the same lensing model, ruling out the strong lensing hypothesis. Both quasar host galaxies are massive (with $M_\text{dyn}\gtrsim10^{10}M_\odot$) and actively forming stars (with $\text{SFR}\gtrsim500M_\odot\text{ yr}^{-1}$). 

The confirmation of J2037--4537 puts a lower limit on the quasar pair fraction at $5.5<z<6$, $F_\text{pair}\geq1.2\%$. This value is significantly higher than the quasar pair fraction at cosmic noon, which is about $10^{-3}$. The elevated quasar pair fraction at high redshifts is also in line with the high GW signal recently reported by PTA experiments.

The confirmation of J2037--4537 enables new insights into the merger-driven quasar triggering in the early Universe. Follow-up observations and modeling for this system will put further constraints on the properties of high-redshift galaxy mergers and quasar pairs.

The code used in this work and the data for making the figures {can be found at \url{10.5281/zenodo.19211007}}. The raw ALMA data can be downloaded via the ALMA Science Archive\footnote{https://almascience.nrao.edu/aq/}.

\begin{acknowledgments}
{We thank the referee for the valuable comments.}
We thank John Silverman for kindly sharing the HSC survey quasar pair fraction. We thank James Nightingale for helping us with using \texttt{PyAutoLens}.

\end{acknowledgments}





%
\facilities{ALMA}

\software{astropy \citep{2013A&A...558A..33A,2018AJ....156..123A,2022ApJ...935..167A},  
CASA, PyAutoLens
          }

\appendix

\section{Examples of Pixel-by-pixel Fitting for the {\cii} Spectra} \label{sec:app}

\begin{figure}
    \centering
    \includegraphics[width=1\linewidth]{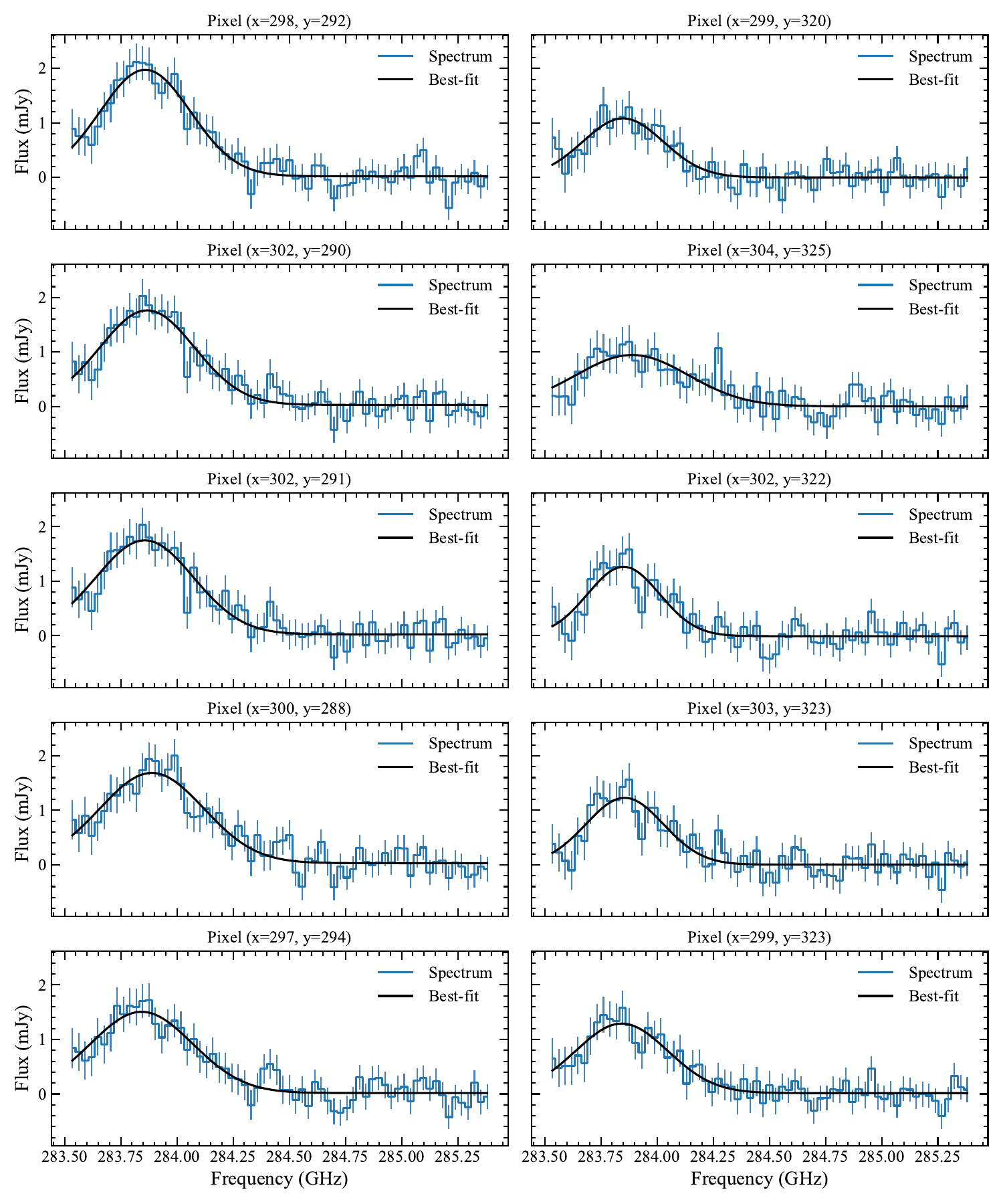}
    \caption{{The {\cii} spectra of example pixels and their best-fit Gaussian emission line model (Section \ref{sec:obs}). For each pixel, we use a Gaussian profile to fit the {\cii} line flux, velocity, and velocity dispersion (Figure \ref{fig:alma}).}}
    \label{fig:pixBpix}
\end{figure}

{Figure \ref{fig:pixBpix} demonstrates example pixel-by-pixel spectra fitting described in Section \ref{sec:obs}. The left column shows pixels from Quasar A, and the right column shows pixels from Quasar B. For individual pixels, a Gaussian profile describes the spectra well.}

\bibliography{sample701}{}
\bibliographystyle{aasjournalv7}



\end{document}